\documentstyle[12pt]{article}
\begin{document}
\textwidth 149mm
\textheight 210mm
\topmargin 0cm
\oddsidemargin 5mm
\newcommand{\eq}[2]{\begin{equation} \label{#1} #2 \end{equation}}
\newcommand{\req}[1]{(\ref{#1})}
\newcommand{\cpn}{$~CP^{n-1}~$}
\newcommand{\topsus}{~topological susceptibility~}
\newcommand{\correl}{~correlation length~}
\hyphenation{per-tur-ba-tive}
\hyphenation{to-po-lo-gi-cal}
\hyphenation{sus-cep-ti-bi-li-ty}
\hyphenation{an-ti-in-stan-ton}
%
\hbox{}
\noindent May 1992 \hfill HU Berlin-IEP-92/2
\begin{center}
\vspace*{1.5cm}
\begin{Large}
{\bf The Topological Susceptibility of the Lattice \cpn Model
on the Torus and the Sphere
\footnote[1] {This research project was partially funded by the Department
of Energy under the contract DE-FG05-87ER40319}
\baselineskip15pt }  \\
\end{Large}

\vspace*{1.5cm}
{\large
N.~Schultka $\mbox{}^*$,
M.~M\"uller-Preussker $\mbox{}^{**}$
}\\
\vspace*{0.7cm}
{\normalsize
$\mbox{}^*$ {\em Department of Physics \\
The Florida State University \\
Tallahassee, FL 32306, USA}\\
$\mbox{}^{**}$ {\em
Humboldt-Universit\"{a}t zu Berlin, Fachbereich Physik \\
Institut f\"ur Elementarteilchenphysik \\
Invalidenstr. 110, O-1040 Berlin,
Germany}}\\     
\vspace*{2cm}
{\bf Abstract}
\end{center}

The topological vacuum structure of the two-dimensional \cpn model for
$~n = 3,5,7~$ is studied on the lattice. In particular we investigate the
small-volume limit on the torus as well as on the sphere and compare with
continuum results. For $~n \ge 5~$ , where lattice artifacts should be
suppressed, the topological susceptibility shows unexpectedly strong
deviations from asymptotic scaling. On the other hand there is an indication
for a convergence to values
obtained analytically within the limit $~n \rightarrow \infty~$ .

\newpage
{\bf 1. Introduction}
\vspace{.5 cm}

The $~2D~$ \cpn model \cite{eichen} is a convenient play-model for testing
non-perturbative methods aimed for application in asymptotically free
field theories like QCD.
Because of the existence of classical (multi-) instanton solutions
it allows a semi-classical treatment  \cite{beco,quasi}. It
can be considered within approximations based on the $~1/n~$ expansion
\cite{dada,munst} as well as within the lattice approach [6 - 14].

One of the main quantities calculated within these different approaches
is the so-called topological
susceptibility $~\chi_{t}~$, i. e. the zero momentum correlator of the
topological charge density. For quenched QCD the analogous
quantity has an important
physical meaning, since it is directly related to the mass of the $~\eta '~$
meson (solution of the $~U_{A}(1)~$ problem \cite{wive}).

However, in computing $~\chi_{t}~$ one meets serious difficulties
both in the continuum
and on the lattice. In the continuum the semi-classical approach can be
developed in a closed form only for the sectors of exact multi-(anti-)
instanton
amplitudes. The contribution of mixed instanton and antiinstanton
configurations to the corresponding functional integral has to rely on model
assumptions, like the dilute instanton gas or liquid models in QCD.
Only in the small-volume limit, where the one-instanton contribution
dominates, reliable results \cite{schw} were obtained consistent with those of
an $~1/n~$ expansion \cite{munst}. Technically this achievement was possible
by formulating the \cpn model on a sphere $~S^2~$. (The same holds for the
pure $~SU(N_{c})~$ Yang-Mills theory treated on $~S^4~$ \cite{lueya}).
On the other hand, on the lattice there is the danger to encounter lattice
artifacts, which were shown to spoil the continuum limit of the topological
susceptibility at least for $~n \leq 3~$ \cite{belu,disl,disb}.
The analogous problem
occurs in pure $~SU(2)~$ lattice gauge theory when the geometric method to
define the topological charge \cite{phist}
is directly applied to the Monte Carlo
generated quantum fluctuations of the gauge field \cite{disym}.

All these difficulties so far have prevented a real consistency check
between the continuum approaches and lattice computations. Since new techniques
for updating the complex vector fields are available \cite{wolf}
we can study the
lattice \cpn model for higher $~n~$, where lattice artifacts should be
suppressed. Furthermore, we treat the model both on a torus and on a sphere.
The latter case gives us the opportunity to investigate the topological
structure in the small-volume limit with the same boundary conditions
as it was done in the continuum case \cite{schw}.

In our previous papers \cite{wirz} we concentrated on the $~CP^{3}~$ model. We
used a lattice formulation of the model, where the Abelian gauge field
involved was considered as a dynamical one. We found a situation which bears
a great resemblance to that in $~SU(2)~$ lattice gauge theory: With the
geometric method one finds an almost perfect asymptotic scaling behaviour
for the topological susceptibility even on very small lattices, although
there are dislocations expected to spoil the continuum limit. Thus, one is
tempted to ask whether the scaling behaviour  is accidental and could be
disturbed if dislocations and other short range fluctuations are suppressed.

Here we come back to another lattice formulation first considered by Petcher
and
L\"uscher \cite{pel} and apply to it the cluster updating algorithm
\cite{wolf,uwe}. The cases $~n~=~3,5,7~$ will be treated. We shall investigate
the topological properties and estimate the correlation lengths $~\xi ~$ both
for the two-dimensional torus $~T^{2}~$ and for the sphere $~S^{2}~$.
We are going to discuss the behaviour of the topological susceptibility down to
volumes  $~V~\simeq ~\xi ^{2}~$ and to compare our lattice results with those
relying on the $~1/n~$ expansion.

Section 2 will collect all continuum results needed for our discussion. The
lattice formulation on $~T^{2}~$ and $~S^{2}~$, respectively, will be
presented in Chapter 3. Section 4 is devoted to the Monte Carlo results.
The conclusions will be drawn in Section 5.
\vspace{1.0 cm}

\noindent{\bf 2. The \cpn Model}
\vspace*{\baselineskip}\\
The \cpn model on an arbitrarily curved two-dimensional manifold is defined by
the following action \cite{eichen}
\eq{e1}{ S = \frac{n}{2f} \int d^{2}x \sqrt{g} \, g^{\mu\nu} \, D_{\mu}z_{a}
        \: \overline{D_{\nu}z_{a}}\;, }
where $g^{\mu\nu}$ denotes the metric tensor. The complex vector field
$z_{a}(x)$,$\; a=1,...,n$
is subjected to the normalization condition
\eq{e2}{z_{a}\:\overline{z}_{a}=1\;. }
$D_{\mu}$ stands for the total covariant derivative, i.e. it consists of
the derivative $\nabla_{\mu}$ in a curved space and in our case a $U(1)$
gauge field $A_{\mu}$,
\eq{e3}{D_{\mu}=\nabla_{\mu}+iA_{\mu}\;. }
$\nabla_{\mu}$ acts on an arbitrary vector $~B_{\mu}~$
in the curved space according to
\eq{na}{\nabla_{\mu} B_{\nu} = \partial_{\mu} B_{\nu} -
                              \Gamma_{\mu\nu}^{\alpha} B_{\alpha} \;,}
where $\Gamma_{\mu\nu}^{\alpha}$ denote the Christoffel symbols.\\
Expressing the gauge field $A_{\mu}$ completely in terms of the $z$-field,
\eq{e5}{A_{\mu}=-\frac{i}{2}\left( z_{a}\partial_{\mu}\overline{z}_{a}
                            - \overline{z}_{a}\partial_{\mu}z_{a} \right),}
enables us to write
\eq{e6}{S = \frac{n}{2f} \int d^{2}x \sqrt{g} \, g^{\mu\nu}
     \left(\; \partial_{\mu}z\partial_{\nu}\overline{z}+
     (z\partial_{\mu}\overline{z})(z\partial_{\nu}\overline{z}) \; \right)\;. }
It is
convenient to introduce projection matrices as follows \cite{pel}
\eq{17}{P_{\alpha\beta}(x)=z_{\alpha}(x) \, \overline{z}_{\beta}(x)}
satisfying
\eq{18}{P=P^{\dag}, \quad P^{2}=P ,\quad \mbox{Tr}P=1 \quad\quad
\forall \, x \;.}
Expressing \req{e1} in terms of \req{17}, we get
\eq{19}{S=\frac{n}{4f} \int d^{2}x\sqrt{g} \, g^{\mu\nu} \, \mbox{Tr}
          \left[ \partial_{\mu} P \partial_{\nu} P \right]\;. }
For our model the topological charge $Q$ can be defined as
\begin{eqnarray} \label{10}
Q &=& \int d^{2}x \, \rho (x) \;, \\
\rho (x) &=& \frac{1}{2\pi}
 \, \varepsilon_{\mu\nu} \, \nabla_{\mu} A_{\nu}\;, \nonumber
\end{eqnarray}
where $\varepsilon_{12}=-\varepsilon_{21}=1$ .
This charge can be understood as  the quantized "magnetic flux"
through the surface. Expressing this "flux" in terms of the
$z$-fields we get
\eq{11}{Q=- \frac{i}{2\pi} \int d^{2}x \, \varepsilon_{\mu\nu} \,
          \partial_{\mu}z \: \partial_{\nu}\overline{z}\;. }
The following relation connects the action \req{e1} and the
topological charge \req{10}
\eq{15}{ S \geq \frac{\pi n}{f} | Q | \;. }
In case that \req{15} becomes an equality, we are led to the self-duality
relation
\eq{16}{ D^{\nu}z=\mp \frac{i}{\sqrt{g}}\varepsilon_{\nu\alpha} \: D_{\alpha}z
\;.}
For the 2-dimensional flat infinite space the (multi-) instanton solutions
of \req{16} look as follows \cite{belu}
\begin{eqnarray} \label{i2}
           z_{\alpha} &=& \frac{p_{\alpha}(x)}{| p(x) |} \;, \nonumber \\
p_{\alpha}&=&C_{\alpha} \prod_{j=1}^{Q}(x_{1}-ix_{2}-a_{\alpha}^{j}),\quad\quad
             C_{n}=1 \;.
\end{eqnarray}
All the $a_{\alpha}^{j}$ are different from each other, and $x_{1},x_{2}$
denote
the coordinates in flat space. Note that the topological charge is
determined by the zeros of the polynomials $p_{\alpha}$. In the case of
our manifold being the sphere $S^{2}$ the coordinates $x_{1},x_{2}$ have to be
understood as the stereographic coordinates.\par
The model is quantized using the functional integration method. Quantum
expectation values of operators $O(z,\overline{z})$ have to be
calculated as follows.
\begin{eqnarray} \label{qu1}
\langle O(z,\overline{z}) \rangle & = & Z^{-1} \int Dz D\overline{z}\,
          O(z,\overline{z})\, \exp( -S ) \;,  \\
 Dz D\overline{z} & = & \prod_{x}\prod_{\alpha=1}^{n} \frac{dz_{\alpha}(x)
        d\overline{z}_{\alpha}(x)}{2\pi i} \;, \quad
\delta \left( \sum_{\beta=1}^{n} z_{\beta}(x)\overline{z}_{\beta}(x)-1 \right)
\end{eqnarray}
with $~Z = \int \, Dz D\overline{z} \, \exp( -S )~$.
The observables we are really interested in throughout this paper are the
\topsus $\chi_{t}$ and the \correl $\xi$. The latter can be read off from the
long-distance behaviour (cf. the next paragraph) of the correlation function
\begin{eqnarray}
C(x,y)&=&\left\langle \left|\sum_{\alpha}z_{\alpha}(x)\overline{z}_{\alpha}(y)
       \right|^{2} \right\rangle - \sum_{\alpha,\beta}
       \left\langle z_{\alpha}(x)\overline{z}_{\beta}(y) \right\rangle
       \left\langle z_{\beta}(y)\overline{z}_{\alpha}(x) \right\rangle
       \nonumber \\  \label{cf}
     &=&\left\langle \left|\sum_{\alpha}z_{\alpha}(x)\overline{z}_{\alpha}(y)
       \right|^{2} \right\rangle - \frac{1}{n}
\end{eqnarray}
The \topsus is defined as follows.
\eq{ap1}{\chi_{t}=\int d^{2}x \langle \rho(x) \, \rho(0) \rangle =
\frac{\langle Q^{2} \rangle}{V} \;, }
where $V$ is the "volume" (i. e. the surface area) of the manifold under
consideration (either the
sphere $S^{2}$ or the torus $T^{2}$). $\chi_{t}$ can be rewritten as
\eq{ap2}{\chi_{t}=\frac{1}{V} \sum_{Q=-\infty}^{Q=\infty} Q^{2} P_{Q}\;, }
with $P_{Q}$ the probability for a field configuration to have  the
topological charge $Q$. On the sphere $S^{2}$ this formula can actually be
exploited within the limit of a small volume \cite{schw} where it  gives
\eq{ap3}{\chi_{t} \simeq \frac{2}{V} P_{1}\;, }
i.e., only the one-(anti-) instanton contribution turns out to be dominant.
The result is
\eq{app4}{ \chi_{t}(V) \Lambda_{MS}^{-2} =
         \frac{2 C_{1}(n)}{V\Lambda_{MS}^{2}}\left[ -\left(
     \frac{V \Lambda_{MS}^{2}}{4\pi} \right)^{\frac{1}{2}} \, \frac{n}{\pi} \,
         \ln \frac{V \Lambda_{MS}^{2}}{4\pi} \right]^{n} \;. }
The coefficients $~C_{1}(n)~$ are numerically known, in particular
\begin{eqnarray}     \label{ap4}
    C_{1}(3) &=& 27.264  \\
    C_{1}(5) &=& \mbox{ }  3.963 \nonumber \\
    C_{1}(7) &=& \mbox{ }  0.343 \nonumber \;.
\end{eqnarray}
On the other hand, the model is $1/n$-expandable \cite{dada}. The result
within the leading order and for $V=\infty$ is
\eq{n1}{ \chi_{t}(\infty) = \frac{3}{4\pi n \xi^{2}} +O\left( \frac{1}{n^{2}}
  \right), }
where $\xi$ denotes the \correl ,  which tends to a finite large $~n~$ limit
\eq{n2}{\xi \Lambda_{MS} =\frac{1}{4\sqrt{\pi}} \mbox{e}^{-\frac{1}{2}
                            \Gamma'(1) } =0.188 \;. }
On the contrary, if the model is treated within the $~1/n~$ expansion
on a sphere $S^{2}$ having
a small finite surface area $~V~$, then $\chi_{t}(V)$ behaves
in accordance with the semi-classical result \req{app4}
\cite{munst}. The limits $n \rightarrow \infty$ and
$V \rightarrow \infty$ do not commute. \par
It is worthwhile for our further considerations
to replace $\Lambda_{MS}$ in \req{app4} by $\xi$ obtained at $V=\infty$
\req{n2}.
\eq{n3}{  \chi_{t}(V) \xi^{2}  =
         \frac{2 C_{1}(n)}{V\xi^{-2}}\left[ -\left(
 \frac{V}{4\pi} \left(\frac{0.188}{\xi} \right)^{2} \right)^{\frac{1}{2}} \,
  \frac{n}{\pi} \,
   \ln \frac{V}{4\pi}\left(\frac{0.188}{\xi}\right)^{2} \right]^{n}  }
\vspace{2\baselineskip} \\
{\bf 3. The  Lattice Formulation}
\vspace{\baselineskip} \\
First, let us describe the lattice constructions for the torus and the sphere.
We decided to use simplicial lattices as the most convenient way
for discretizing curved manifolds. In the torus case it will allow us directly
to compare with \cite{pel} and, as we shall explain below, with $~1/n~$
expansion results.

\paragraph{Torus $T^{2}$:} The plaquettes of the lattice are equilateral
triangles with edge length $a$. A lattice point $\vec{x}$ is then described
by
\begin{center} $ \vec{x} =x_{1}\vec{n}_{1}+x_{2}\vec{n}_{2} $ \end{center}
where
\begin{eqnarray} \label{v1}
    \vec{n}_{1} &=& a \; (1,0),  \\ \label{v2}
    \vec{n}_{2} &=& a \left( \frac{1}{2}, \frac{\sqrt{3}}{2} \right)
\end{eqnarray}
The topology of the torus is established by imposing periodic boundary
conditions on the functions defined on the lattice.
\eq{39}{f\left( \vec{x}+N_{1}\vec{n}_{1}+N_{2}\left(\vec{n}_{2}-\vec{n}_{1}
         \right)\right)=f\left(\vec{x}\right) }
where $N_{1}$ and $N_{2}/2 $ are integers.
The surface area $V$ of the torus is
$$ V=\frac{\sqrt{3}}{2}N_{1}N_{2}a^{2}. $$
\paragraph{Sphere $S^{2}$:}
The sphere $~S^2~$ is discretized by a simplicial lattice in the following
way. We start from a regular tetrahedron with corners
placed at the surface of the
sphere. Then we make several {\it steps of refinement} of this initial lattice.
During each step we project the center of each given link
onto the surface to introduce further sites. All sites, the old and new ones,
are connected with their nearest neighbours by straight lines. After each step
every simplex is replaced by four new simplices. After $~N~$ steps we have
$~P = 2(4^{N}+1)~$ sites and $~L = 3(P-2)~$ links.
Figs. 1, 2 show subsequent steps of refinement and the obtained link length
distribution, respectively.
In order to define a metric on these lattices, we consider the triangles to be
flat. Identifying one of the  \mbox{corners ~i~} with the origin
of the local coordinate system, we have for the triangle \mbox{(i,j,k)}
\cite{itb}
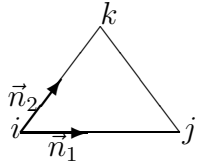
\begin{center}
\begin{picture}(60,70)
\put(5,10){\line(1,0){60}}
\put(5,10){\line(3,4){30}}
\put(35,50){\line(3,-4){30}} \thicklines
\put(5,10){\vector(1,0){25}}
\put(5,10){\vector(3,4){15}} \thinlines
\put(1,8){$i$} \put(35,51){$k$} \put(66,8){$j$}
\put(15,1){$\vec{n}_{1}$}   \put(0,20){$\vec{n}_{2}$}
\end{picture}
\end{center}
the following metric tensor

\begin{eqnarray}
g^{\mu\nu} &=& \frac{l^{2}_{1}l^{2}_{2}}{4\bigtriangleup^{2}}M^{\mu\nu}
       \nonumber \\  \label{22}
M &=&  \left( \begin{array}{cc}
        1                        &          -\vec{n}_{1} \vec{n}_2 \\
       -\vec{n}_{1} \vec{n}_2    &          1
       \end{array} \right)\;,
\end{eqnarray}

where $~\bigtriangleup~$ is the area of the triangle and $~l_{i}~$
are the lengths of the links corresponding to the unit vectors
$\vec{n}_{i}$.\par
\paragraph{Lattice action:}
The most convenient way to discretize the lattice action uses its
representation
in terms of the projection matrices \req{19}. Thus we have
\eq{20}{S=\frac{n}{4f} \sum_{\bigtriangleup} \, \bigtriangleup \,
     g^{\mu\nu} \, \mbox{Tr}\left[\partial_{\mu} P \partial_{\nu} P \right]
\;,}
where the partial derivative has to be taken as follows:
\eq{23}{\partial_{\mu}P=\frac{P(x+l_{\mu} \vec{n}_{\mu})-P(x)}{l_{\mu}}
        \quad\quad  \mu=1,2 \;. }
\req{20} can be rewritten as a sum over links $~\langle ij \rangle~$ between
nearest neighbour sites $~i,j~$
\eq{26}{S= \beta \sum_{\langle ij \rangle} S_{ij} \quad , \quad
        \beta=\frac{n}{4f} \; }
with
\eq{27}{S_{ij}=\Gamma_{ij}\, \left( \, 1- |z(i)\overline{z}(j)|^{2} \, \right)}
and the weight factors
\eq{28}{\Gamma_{ij}=\frac{\vec{l}_{kj}\vec{l}_{ki}}{2\bigtriangleup(i,j,k)}
              +\frac{\vec{l}_{k'j}\vec{l}_{k'i}}{2\bigtriangleup(i,j,k')} \;. }
The corresponding geometry is shown in the figure.
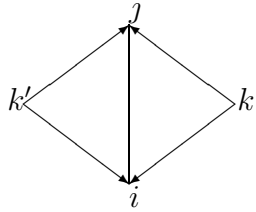
\begin{center}
\begin{picture}(90,70)
\put(5,35){\vector(4,3){40}}
\put(5,35){\vector(4,-3){40}}
\put(85,35){\vector(-4,3){40}}
\put(85,35){\vector(-4,-3){40}}
\put(45,5){\line(0,1){60}}
\put(-1,33){$k'$} \put(86,33){$k$} \put(45,68){$j$} \put(45,-4){$i$}
\end{picture}
\end{center}
The action \req{26} is obviously $U(1)$ gauge invariant. Note that in the
case of the torus $T^{2}$ the weights are constant:
\eq{29}{\Gamma_{ij}=\frac{2}{\sqrt{3}} ,}
i.e., we obtain the action invented in \cite{pel}.
\vspace{\baselineskip} \par
\paragraph{Topological charge:}
For a given field configuration on the lattice, the topological charge
is computed by using its definition according to
\cite{pel}.
\eq{30}{Q=\sum_{\bigtriangleup} q(\bigtriangleup) \quad \quad
             q(\bigtriangleup)\, \epsilon \, (-\frac{1}{2},\frac{1}{2}) \;, }
where
\eq{31}{\exp(i2\pi q(\bigtriangleup) ) =
       \frac{\mbox{Tr} \left(\, P(k)P(j)P(i) \, \right)}{\left|
       \mbox{Tr}( \, P(k)P(j)P(i) \,) \right| } \;. }
The sum runs over all (positively oriented) triangles. Equation
\req{31} rewritten in terms of the $z$-field means
\begin{eqnarray} \label{32}
\exp(i2\pi q(\bigtriangleup) ) &=&
             \frac{z(i)\overline{z}(j)}{\left|z(i)\overline{z}(j)\right|} \,
             \frac{z(j)\overline{z}(k)}{\left|z(j)\overline{z}(k)\right|} \,
             \frac{z(k)\overline{z}(i)}{\left|z(k)\overline{z}(i)\right|}
        \\
   &=& \exp(i(\, \phi(i,j)+\phi(j,k)+\phi(k,i)\,))
\end{eqnarray}
and finally
\eq{34}{2\pi q(\bigtriangleup)=\phi(i,j)+\phi(j,k)+\phi(k,i)+
        2\pi m(\bigtriangleup)  \quad\quad \epsilon \, (-\pi,\pi], }
with $m(\bigtriangleup)$ a properly chosen integer.\par
\paragraph{Monte Carlo simulation:}
The expectation values \req{qu1} are computed numerically using the
Monte Carlo method. We generate equilibrium field configurations by applying
the one-cluster algorithm invented by U.Wolff \cite{wolf}.
This algorithm can be easily implemented for arbitrary $~n~$ and works
reasonably fast, in spite of the fact that it does not remove critical
slowing down \cite{uwe}.
We choose an arbitrary lattice site $~x_{0}~$
and collect adjacent lattice points into the cluster with the bond probability
\eq{46}{p=(1-\exp(-\beta \triangle S_{ij}))\,
                      \theta ( \triangle S_{ij}) \;, }
\eq{47}{\triangle S_{ij} = S_{ij}\left(\, R\, z(i),z(j) \, \right) -
                           S_{ij}\left(\, z(i),z(j) \, \right) \;,}
with  $S_{ij}$ the link action \req{27} and
\eq{48}{\theta (\triangle S_{ij}) = \left\{
             \begin{array}{ll}
                     1 & \mbox{for} \: \triangle S_{ij}>0 \\
                     0 & \mbox{otherwise}
             \end{array}\right. }
Having found all sites
belonging to the cluster, the field vectors sitting at these sites
are flipped. The flip operation $~R~$ acting on the fields $~z~$ is defined
as follows:
\eq{43}{R\, z=z-2(\overline{r}z)r, }
where $r$ is a random vector in the $z$-space, i.e.,
\eq{al1}{ \sum_{\alpha=1}^{n} r_{\alpha} \overline{r}_{\alpha} = 1. }
Note that
\eq{44}{R^{2}\, z=z \;, }
\eq{45}{\left( R\, z(i) \right)\overline{\left( R\, z(j) \right)}=
               z(i)\overline{z(j)} \;. }
Choosing the bond probability and the flip operation as above endows
the algorithm with detailled balance \cite{wolf,bind}.
Fig. 3  shows the cluster point distribution for the $CP^{6}$ Model on
the torus $T^{2}$ with $\beta=4.6$ on a $100 \times 100$ lattice. The
average number of cluster points is 6516.\par
In order to test the reliability of our program we compared the expectation
value of the link action \req{27} with analytical results.
On the torus, the following weak coupling formula holds \cite{pel}
\eq{51}{\beta \gg 1 \quad \quad
       \langle S_{ij} \rangle=\frac{2}{3}\left( \frac{n-1}{2} \frac{1}{\beta}+
            \frac{(n-1)n}{16\sqrt{3}} \frac{1}{\beta^{2}} +\cdots \right) }
and for $~\beta = 0~$ the strong coupling ones
\eq{52}{\langle S_{ij} \rangle = \frac{2}{\sqrt{3}} \frac{n-1}{n} \quad
       \quad  \mbox{on the torus $T^{2}$, }}
\begin{eqnarray} \label{53}
   \langle S_{ij} \rangle & = & \overline{\Gamma}_{ij} \frac{n-1}{n} \quad
        \quad  \mbox{on the sphere $~S^{2}~$, where} \\
   \overline{\Gamma}_{ij} &=& \frac{1}{L} \sum_{ij} \Gamma_{ij} \quad
   \quad \mbox{with $~L~$ the number of links.}
   \nonumber
\end{eqnarray}
Fig. 4 shows the analytical curve \req{51} compared with our numerical
values for $S_{ij}$ versus $\beta$ for the $CP^{2}$ and $CP^{4}$ Model
on the torus $T^{2}$. For $\beta = 0$, we found
\begin{center}
\begin{tabular}{| *{3}{c|} }
\hline
 & \multicolumn{2}{c|}{$\langle S_{ij} \rangle $} \\
\cline{2-3}
 & Analytic  & Numeric \\
 & (\req{52} resp. \req{53}) & \\
\hline
torus $T^{2}$ & & \\
\cline{1-1}
 $CP^{2}$ Model & .7698     & .7701     \\
 $CP^{4}$ Model & .9238     & .9231     \\
\hline \hline
sphere $S^{2}$ & & \\
$\overline{\Gamma}_{ij}=1.5416 \; (N=3) $ & & \\
\cline{1-1}
 $CP^{2}$ Model & 1.0277     & 1.0266     \\
 $CP^{4}$ Model & 1.2333     & 1.2318     \\
\hline
\end{tabular}
\end{center}
The purely statistical errors are omitted because they were found of order
$10^{-5}$.
\medskip

In order to compute the correlation function, we used the
{\it improved estimator method} \cite{wolf2,uwe}, i.e.,
\begin{eqnarray}
C(x)& = & (n+1)\left\langle \frac{P}{|c|} \theta(x,c) \theta(0,c)
            \mbox{Re}\left[ \, ( \overline{z}(x)z(0) \right. \right.
            \nonumber \\  \label{54}
& & \mbox{} - \left. \left.
            (\overline{z}(x)r)(\overline{r}z(0)) )
            \overline{(\overline{z}(x)r)(\overline{r}z(0))}\, \right]
            \right\rangle
\end{eqnarray}
where $r$ denotes the  reflexion vector \req{43} and
\eq{55}{\theta(x,c)=\left\{
        \begin{array}{ll}
          1 & \mbox{if $x$ belongs to the cluster }\\
          0 & \mbox{otherwise.}
        \end{array}  \right. }
$P$ and $|c|$ are the number of all lattice points and  the number of all
cluster points respectively.
\paragraph{Correlation lengths:} $~\xi~$ is extracted as follows.
\paragraph{Torus $T^{2}$:} Representing the point $x$ as
$\vec{x}=x_{1}\vec{n}_{1}+x_{2}(\vec{n}_{2}-\vec{n}_{1})$ and
using the vectors $\vec{n}_{1}$ and $\vec{n}_{2}$ from \req{v1} and \req{v2},
we actually compute
\eq{56}{ \widetilde{C}(x_{2})= \sum_{x_{1}=0}^{l_{1}-1} C(x_{1},x_{2}). }
In the range $ 0 \ll x_{2} \ll l_{2} $, we expect the following behaviour
\eq{57}{\widetilde{C}(x_{2})=A \cosh \frac{x_{2}-\frac{l_{2}}{2}}{\xi} ,}
where the constant $A$ and the \correl $\xi$ are extracted by fitting
the Monte Carlo data.
\paragraph{Sphere $S^{2}$:} In order to find the shape of the correlation
function we start from a free, scalar, massive, Euclidean theory on
the sphere $S^{2}$. The propagator is as usual
\eq{c1}{G=\frac{1}{\triangle+m^{2}} ,}
\eq{c2}{\triangle=-\frac{1}{\sqrt{g}}\partial_{\mu}\left(
\sqrt{g}\partial^{\mu}
                    \right). }
Using spherical coordinates, the following equation has to be solved:
\eq{c4}{\left(
-\frac{1}{R^{2}}\frac{1}{\sin^{2}\vartheta}\partial^{2}_{\varphi}
               -\frac{1}{R^{2}\sin\vartheta}\partial_{\vartheta}( \sin\vartheta
                \partial_{\vartheta} ) + m^{2} \right)
 G(\varphi,\vartheta,\varphi',\vartheta')=\delta(\varphi,\vartheta,\varphi',
               \vartheta') }
or
\begin{eqnarray}
\left( -\frac{1}{\sin^{2}\vartheta} \partial^{2}_{\varphi} -
     \frac{1}{\sin\vartheta}\partial_{\vartheta}(\sin \vartheta
      \partial_{\vartheta}) +
 (m R)^{2} \right)& & \!\!\!\!\!\!\!\!\!\!
  G(\varphi,\vartheta,\varphi',\vartheta')= \nonumber \\
    \label{c10}
  & & \!\!\!\!\!\!\!\!\!\!  \sum_{l,m}
          Y_{l,m}(\varphi,\vartheta) \overline{Y}_{l,m}(\varphi',\vartheta'),
\end{eqnarray}
where the bar means complex conjugation. Thus,
\eq{c13}{G(\varphi,\vartheta,\varphi',\vartheta')=
       \frac{1}{4\pi} \sum_{l=0}^{\infty}
       \frac{2l+1}{l(l+1) + (m R)^{2} } P_{l}(\cos \widetilde{\vartheta}) ,}
where $\widetilde{\vartheta} \epsilon [0,\pi]$ denotes the angle between
the points $(\varphi,\vartheta)$ and $(\varphi',\vartheta')$. We fitted
the Monte Carlo data according to
\eq{58}{ C(\vartheta(x),0)= A \sum_{l=0}^{\l_{max}} \frac{2l+1}{l(l+1)+
      \left( \frac{R}{\xi} \right)^{2} } P_{l}(\cos \vartheta)  }
in the range $ 0 \ll \vartheta(x) \leq \pi $, truncating the series
at $l_{max}=100$, which provides stable results.
\medskip
\paragraph{Scaling:}
Physical observables, e. g. $~\chi_{t}= \langle Q^{2} \rangle /V~$ and $~\xi~$,
are measured in units of the average link length
$\overline{l}$, which is related to $~\beta~$ through the 2~-~loop
renormalization group expression \cite{pel}
\eq{ren2}{\Lambda_{L} \overline{l}= \left( \frac{4\pi}{n}\beta \right)^{
           \frac{2}{n} } \mbox{e}^{-\frac{4\pi}{n} \beta}. }
Then, approaching the continuum limit $\chi_{t}\overline{l}^{2}$
and $\xi\overline{l}^{-1}$ are expected to behave like
\begin{eqnarray}\label{ren33}
 \xi \overline{l}^{-1} &=& C_{\xi}\left( \frac{4\pi}{n}\beta \right)^{-
           \frac{2}{n} } \mbox{e}^{\frac{4\pi}{n} \beta}\\ \label{ren44}
 \chi_{t}\overline{l}^{2} &=& C_{\chi}\left( \frac{4\pi}{n}\beta \right)^{
           \frac{4}{n} } \mbox{e}^{-\frac{8\pi}{n} \beta}
\end{eqnarray}
(On $~T^{2}~$ replace $~\overline{l}~$ by the lattice spacing $~a~$.)

We have checked the occurence of dislocations
(i.e. lattice fields $~z_{\alpha}~$
that have a topological charge $Q=\pm 1$ with respect to definition \req{30},
\req{31} but have an action below $~8 \pi \beta /n~$ i. e. the scaling
exponent of $\chi_{t} \overline{l}^{2}$) by minimizing the action of the
$CP^{1}$ model while keeping
the topological charge fixed. We have found
\begin{center}
\begin{tabular}{|*{2}{c|}}
\hline
 & $S_{d}$ \\
\hline
 sphere $S^{2}$  & $2.09\pi\beta$ \\
\hline
 torus $T^{2}$  & $2.06\pi\beta$ \\
\hline
\end{tabular}
\end{center}
For higher values $~n~$ the $~CP^{1}~$ configurations can be simply embedded.
Therefore the dislocations found become suppressed for sufficiently large
$~\beta~$ at $~n \ge 4~$. \par
Let us consider the lattice \cpn Model for large $n$ on the torus $T^{2}$.
Then \req{ren2} enables us to find the curve that $\xi a^{-1}$ and
$\chi_{t} a^{2}$ should lie on if we approach the infinite volume limit.
Equations \req{n1} and \req{n2} provide us with
\eq{res1}{ \chi_{t}(\infty)a^{2}=\frac{12}{n}\mbox{e}^{\Gamma'(1)}\,
    (a\Lambda_{L})^{2}\, \left(
\frac{\Lambda_{MS}}{\Lambda_{L}}\right)^{2} .}
Introducing \req{ren2} and
\eq{res2}{ \frac{\Lambda_{MS}}{\Lambda_{L}} = 2\sqrt{3} \exp \left(
      -\frac{1}{2} (\ln\pi +\Gamma'(1)) + \frac{\pi}{\sqrt{3}} \right) }
\cite{pel} into \req{res1} gives $\chi_{t}(\infty) a^{2}$.
For $~n \ge 4~$ we expect
our data to approach this curve for increasing $n$ and $\xi \gg a$.
For the \correl, we get
\eq{rez1}{\xi a^{-1}= \frac{1}{4\sqrt{\pi}}\mbox{e}^{-\frac{1}{2}\Gamma'(1)}
       \frac{1}{a\Lambda_{L}}\frac{\Lambda_{L}}{\Lambda_{MS}}.  }
\vspace{2\baselineskip} \\
{\bf 4. Results}
\vspace{\baselineskip} \\
We have simulated the \cpn model with $~n~=~3~$, 5, and 7,
each case at a variety
of $~\beta~$ values and lattice sizes up to $~10^{2}~$ and $~60*120~$,
respectively,
for $~T^{2}~$ and up to division $~N~=~6~$ for $~S^{2}~$.
Typically we carried out $~10^{4}~$
cluster updates for thermalization followed by $~O(10^{3})~$ $(~O(10^{4})~)$
measurements of the topological charge (correlation function) separated by
$~10~$ $(~5~)$ cluster updates. A cluster update means building up one cluster
and flipping all the $~z's~$ within this cluster.

Figs. 5 - 9 show our main data. The errors of the topological susceptibility
indicated were estimated by collecting the measured values into groups of
$~O(100)~$ single measurements and by taking the statistical error of the
group averages. In order to determine the correlation length we splitted our
data for the correlation function into $~5~$ groups. For each of them we fitted
the correlation length separatly and estimated the error from these values.

In Figs. 5 we show the topological susceptibility for the cases $~n=3,~5,~$
and 7 on $~T^{2}~$ as a function of the "$~n~$-invariant" coupling
$~f^{-1}~=~4 \pi \beta/n~$. The approximate scaling behaviour or weak scaling
violation seen for $~CP^{2}~$ (Fig. 5a) is in accordance with the observations
in \cite{pel} and is related to the dominance of dislocations. In fact, at
larger $~f^{-1}~$ values the Monte Carlo time-histories for the topological
charges have shown us the dominance of short-range fluctuations (i. e. only
very
short appearance of charges being unequal to zero).

For the cases $~CP^{4}~$ and $~CP^{6}~$, where dislocations become definitely
suppressed, we see a surprisingly strong deviation from asymptotic scaling in
the opposite direction. This is in contrast to the conclusions drawn in
\cite{uwe} and in agreement with those of \cite{hasmey}. The investigation
in \cite{uwe} concentrated on data produced in a smaller $~f^{-1}~$-range.
Thus the falling tail at smaller bare coupling was missed. Nevertheless,
the data presented there (with another lattice discretization) are compatible
with ours. The time histories of topological charges show freezing effects
within sectors of non-vanishing $~Q~$ over a few hundreds of consecutive
charge measurements. This means long-range topological fluctuations
occur. In spite of this, within the given $~\beta~$ - or $~f^{-1}~$ -range the
cluster algorithm allows for sufficient tunneling between the sectors such that
we consider our results for the \topsus reliable up to the largest $~f^{-1}~$
-values shown.

In Figs. 5b,c, for comparison, we have drawn also the leading $~1/n~$ result
for $~\chi_{t}(\infty)~$ according to \req{res1}. There seems to be a tendency
to reach the $~1/n~$ and continuum limit at very large $~\beta~$ !

Figs. 6 present the analogous data for the case of the sphere $~S^{2}~$. The
observations and conclusions concerning the scaling violation are definitely
the same. The $~1/n~$ limit cannot be given here because of the unknown
$~\Lambda~$-ratio (compare with \req{res2}). The comparison of Figs. 5 with
Figs. 6 shows stronger finite-size effects in the $~S^{2}~$-case than for the
torus $~T^{2}~$.

In order to study the behaviour of the \topsus as a function of the physical
volume we measured the \correl $~\xi~$. Figs. 7, 8 show the results of our
fits to the correlation functions as described above and represented as
functions of $~f^{-1}~$. Figs. 7a, b present the cases $~CP^{4}~$ and
$~CP^{6}~$ on $~T^{2}~$, whereas Fig. 8 shows $~CP^{6}~$ on $~S^{2}~$.
All these data obtained for the \correl are compatible with asymptotic scaling
or even with slight scaling violations as reported in \cite{hasmey}. About
possible convergence to the $~1/n~$-limit we cannot draw here any conclusion.
We need better statistics by at least one order of magnitude or/and an improved
cluster, overrelaxation or multigrid algorithm \cite{hasmey}.

Finally, in Fig. 9 we study the volume dependence of the \topsus for $~n~=~7~$
and $~S^{2}~$ taking $~\chi_{t}~$ and the volume $~V~$
(surface area of $~S^{2}~$) in units of the \correl $~\xi~$, which itself
corresponds to the infinite-volume limit. The big error bars are due to the
statistical errors of $~\xi~$. Nevertheless, we see the expected tendency
of a suppression of topological fluctuations in the limit of a small volume
$~V~$. Fig. 9 contains the analytic result \req{n3}, too, represented by a
continuous line. Thus, we are qualitatively in the right ball park. But a real
quantitative comparison with the continuum behaviour below
$~V~ \simeq ~\xi^{2}~$ needs yet large efforts in future.
\vspace{2\baselineskip} \\
{\bf 5. Conclusions}
\vspace{\baselineskip} \\
In this paper we have studied the topological properties of the $~2D~$
\cpn \\
model on a lattice for varying $~n~$ and for the cases of a (flat) torus
and a sphere. We used the cluster update algorithm which can be economically
implemented, and which allows to treat larger $~n~$'s and larger lattice sizes
at intermediate coupling values than the standard Metropolis algorithm. We
convinced ourselves that there is a reasonable tunneling rate between different
topological sectors with long-living, large-scale field configurations
in the coupling range considered. This does not mean
that the algorithm allows to suppress critical slowing down.
The contrary was shown in \cite{uwe}.

The main quantity considered was the \topsus . It shows a strong violation
effect from asymptotic scaling for $~n \geq 5~$ on $~T^{2}~$ as well as on
$~S^{2}~$. In this case dislocations are known to become suppressed for
sufficiently large $~\beta~$. On the other hand for $~n=3~$ we recovered
approximate scaling or slight scaling violation into the opposite direction
due to the presence of dislocations.

We compared the outcome of the Monte Carlo investigation with the analytical
leading order $~1/n~$ result for infinite volume. Our data provide an
indication that for very large $~\beta~$ the $~1/n~$ result combined with
asymptotic scaling could be reached.

Unfortunately, within the given statistics
the improved estimator method did not provide us correlation
lengths $~\xi~$ for large volumina with sufficient high accuracy in order to
study the limit of small volume $~V \rightarrow 0~$ in terms of physical units
$~\xi~$. So the comparison of the semi-classical and $~1/n~$ behaviour
on the sphere $~S^{2}~$ in this limit deserves further investigations. So far
we obtained qualitatively a reasonable behaviour at $~V \geq \xi^{2}~$.
The multigrid method can provide a possibility to improve that. We hope to come
back to this question very soon.

\vspace*{1cm}
\noindent {\large \bf Acknowledgements}

We would like to thank M. L\"uscher, U.-J. Wiese for valueable
discussions.
Also one of us (M.M.-P.) would like to express
his deep gratitude to all collegues at Bielefeld University for
the kind hospitality extended to him. We are indebted to M. Hasenbusch and
St. Meyer for sending us their data for $~CP^{3}~$ prior to publication.

\newpage
\noindent {\large \bf Figure captions}

\vspace{.5cm}
\noindent {\bf Fig.~1:~} First three steps of refinement for discretizing the
sphere $~S^{2}~$.

\vspace{.5cm}
\noindent {\bf Fig.~2:~} Link length distribution on $~S^{2}~$
for refinement step
$~N~=~5~$. Average link length is $~\overline{l}~=~.08794~$ for a unit sphere.

\vspace{.5cm}
\noindent {\bf Fig.~3:~} Distribution of cluster sizes produced with the
one-cluster algorithm \cite{wolf} for $~T^{2}~$ with lattice size
$~100 \times 100~$ and $~CP^{6}~$ at $~\beta = 4.6~$. The average size is 6516.

\vspace{.5cm}
\noindent {\bf Fig.~4:~} Comparison of numerical $~<~S_{ij}~>~$ values with
analytic ones (continuous lines) at weak coupling. Squares (triangles)
correspond to $~CP^{2}~$ ($~CP^{4}~$).

\vspace{.5cm}
\noindent {\bf Fig.~5:~} Topological susceptibility as a function of $~f^{-1}~$
for the torus $~T^{2}~$ and
$~CP^{2}~$ ({\bf a}),
$~CP^{4}~$ ({\bf b}) and
$~CP^{6}~$ ({\bf c}), respectively.
The straight line corresponds to an asymptotic scaling behaviour. The dashed
line is the $~1/n~$ result according to \req{res1}.

\vspace{.5cm}
\noindent {\bf Fig.~6:~} Topological susceptibility as a function of $~f^{-1}~$
for the sphere $~S^{2}~$ and
$~CP^{2}~$ ({\bf a}),
$~CP^{4}~$ ({\bf b}) and
$~CP^{6}~$ ({\bf c}), respectively.
The straight line corresponds to asymptotic scaling.

\vspace{.5cm}
\noindent {\bf Fig.~7:~} Correlation length $~\xi~$ as a function of $~f^{-1}~$
for $~T^{2}~$ and
$~CP^{4}~$ ({\bf a}) and
$~CP^{6}~$ ({\bf b}), respectively.
Straight and dashed lines as in Figs. 5.

\vspace{.5cm}
\noindent {\bf Fig.~8:~} Correlation length $~\xi~$ as a function of $~f^{-1}~$
for $~S^{2}~$ and $~CP^{6}~$.

\vspace{.5cm}
\noindent {\bf Fig.~9:~} Topological susceptibility as a function of the
physical volume, both expressed in units of the correlation length. The
continuous line shows the one-instanton result \req{n3} acc. to \cite{schw}.
The dashed line indicates the infinite volume $~1/n~$ result, cf. Eq.
\req{n1}.

\end{document}